\documentclass[
showpacs,
floatfix,
aps,
prb,
amsmath,
twocolumn,
superscriptaddress, 
tightenlines
]{revtex4-1}

\usepackage{graphicx}
\usepackage{dcolumn}
\usepackage{amsmath}
\usepackage{amsfonts}
\usepackage{bm}
\usepackage{epsfig}
\usepackage{times}
\usepackage[varg]{txfonts}
 
\newcommand{\be}{\begin{equation}}
\newcommand{\ee}{\end{equation}}
\newcommand{\bea}{\begin{eqnarray}}
\newcommand{\eea}{\end{eqnarray}}

\def\eac{\epsilon}

\def\oc{\omega_{\mbox{\scriptsize {c}}}}

\def\tq{\tau_{\mbox{\scriptsize {q}}}}

\newcommand{\req}[1]{Eq.\,(\ref{#1})}

\newcommand{\rfig}[1]{Fig.\,\ref{#1}}
\newcommand{\rFig}[1]{Figure\,\ref{#1}}
\newcommand{\rref}[1]{Ref.\,\onlinecite{#1}}

\begin{document}

\title{Microwave-induced resistance oscillations in tilted magnetic fields}

\author{A. Bogan}
\affiliation{National Research Council of Canada, Ottawa, Ontario K1A 0R6, Canada}
\affiliation{Department of Physics and Astronomy, University of Waterloo, Waterloo, Ontario N2L 3G1, Canada}

\author{A.\,T. Hatke}
\affiliation{School of Physics and Astronomy, University of Minnesota, Minneapolis, Minnesota 55455, USA}

\author{S.\,A. Studenikin}
\affiliation{National Research Council of Canada, Ottawa, Ontario K1A 0R6, Canada}

\author{A. Sachrajda}
\affiliation{National Research Council of Canada, Ottawa, Ontario K1A 0R6, Canada}

\author{M.\,A. Zudov}
\email[Corresponding author: ]{zudov@physics.umn.edu}
\affiliation{School of Physics and Astronomy, University of Minnesota, Minneapolis, Minnesota 55455, USA}

\author{L.\,N. Pfeiffer}
\affiliation{Princeton University, Department of Electrical Engineering, Princeton, New Jersey 08544, USA}

\author{K.\,W. West}
\affiliation{Princeton University, Department of Electrical Engineering, Princeton, New Jersey 08544, USA}


\begin{abstract}
We have studied the effect of an in-plane magnetic field on microwave-induced resistance oscillations in a high mobility two-dimensional electron system.
We have found that the oscillation amplitude decays exponentially with an in-plane component of the magnetic field $B_\parallel$.
While these findings cannot be accounted for by existing theories, our analysis suggests that the decay can be explained by a $B_\parallel$-induced correction to the quantum scattering rate, which is quadratic in $B_\parallel$. 
\end{abstract}
\maketitle

Microwave-induced resistance oscillations (MIRO)\citep{zudov:2001a,ye:2001} and associated zero-resistance states\citep{mani:2002,zudov:2003} are prime examples of nonequilibrium transport phenomena,\citep{zudov:2001b,yang:2002,zhang:2007c,zhang:2008,khodas:2010,wiedmann:2010b,hatke:2010a,dai:2010,hatke:2011b} which occur in high mobility two-dimensional electron systems (2DES) subject to a weak perpendicular magnetic field, $B_\perp$.
Owing to both theoretical\citep{andreev:2003,durst:2003,lei:2003,vavilov:2004,dmitriev:2003,dmitriev:2005,khodas:2008,dmitriev:2009b,hatke:2011e} and experimental\citep{zudov:2004,willett:2004,mani:2004a,mani:2004e,studenikin:2004,studenikin:2005,smet:2005,mani:2005,yang:2006,zudov:2006a,zudov:2006b,studenikin:2007,hatke:2008a,hatke:2008b,hatke:2009a,mani:2010,dorozhkin:2011,hatke:2011c,hatke:2011e,hatke:2011f} progress, our understanding of these phenomena has improved dramatically over the last decade.
Theoretically, MIRO are usually discussed in terms of two distinct mechanisms, referred to as the \emph{displacement}\citep{durst:2003,lei:2003,vavilov:2004} and the \emph{inelastic}.\citep{dmitriev:2005}
In the regime of overlapping Landau levels and linear in microwave power, the theories predicts that high order MIRO can be described by a radiation-induced correction to the resistivity (photoresistivity) of the form\citep{dmitriev:2009b}
\be
\delta \rho_\omega \propto -  \eac \lambda^2 \sin 2\pi\eac\,,
\label{eq.miro}
\ee
where $\eac = \omega/\oc$, $\omega=2\pi f$ and $\oc=eB_\perp/m^\star$ ($m^\star$ is the effective mass of an electron) are the microwave and the cyclotron frequency, respectively, $\lambda =\exp(-\eac/2\eac_0)$ is the Dingle factor, $\eac_0=f\tq^0$, and $\tq^0$ is the quantum lifetime.

Over the past decade, many experiments have examined the functional dependences of the MIRO amplitude on magnetic field,\citep{zudov:2001a,hatke:2009a,hatke:2011b,hatke:2011f}  power,\citep{ye:2001,zudov:2003,studenikin:2004,willett:2004,mani:2004a,mani:2010,hatke:2011e} and temperature,\citep{studenikin:2005,studenikin:2007,hatke:2009a} but unsolved puzzles remain.
One such puzzle is the role of an in-plane magnetic field, $B_\parallel$, which has been investigated in two independent experiments\citep{mani:2005,yang:2006} with conflicting outcomes.
While both studies agreed that the MIRO frequency is governed by the perpendicular component, $B_\perp$, in \rref{mani:2005} MIRO remained essentially unchanged up to $B_\parallel > 10$ kG, while in \rref{yang:2006} MIRO were strongly suppressed by $B_\parallel \simeq 5$ kG.
This controversy, and the lack of explanation of the suppression observed in \rref{yang:2006}, 
indicate that the role of $B_\parallel$ on microwave photoresistance deserves further studies.

In this paper we systematically investigate the effect of an in-plane magnetic field on microwave-induced resistance oscillations in a high mobility 2DES.
We find that with increasing tilt angle $\theta$, MIRO get strongly suppressed by $B_\parallel$ of a few kG.
The observed suppression is nonuniform in a sense that it depends on the oscillation order; with increasing $\theta$, the lower order oscillations decay faster than the higher order oscillations.
We discuss our findings in the context of \req{eq.miro} and show that the suppression can be understood in terms of a $B_\parallel$-induced increase of the quantum scattering rate.
This correction is found to scale with $B_\parallel^2$, but the exact origin of such modification remains a subject of future studies.

Our sample was cleaved from a symmetrically doped GaAs/Al$_{0.24}$Ga$_{0.76}$As 300\,\AA-wide quantum well grown by molecular beam epitaxy.
A Hall bar mesa of a width $w=200$ $\mu$m was fabricated using photolithography. 
Ohmic contacts were made by evaporating Au/Ge/Ni and thermal annealing in forming gas.
After illumination with red light-emitting diode, electron density and mobility were $n_e \approx 3.6 \times 10^{11}$ cm$^{-2}$ and $\mu \approx 1.3 \times 10^7$ cm$^2$/Vs, respectively.
Microwave radiation of frequency $f = 48.4$ GHz was delivered to the sample via a semirigid coaxial cable terminated with a 3 mm antenna.\citep{studenikin:2005,studenikin:2007}
A split-coil superconducting solenoid allowed us to change the magnetic field direction \emph{in situ}, by rotating the $^3$He insert, without disturbing the distribution of the microwave field.
The longitudinal resistivity was recorded using low-frequency (a few hertz) lock-in amplification under continuous microwave irradiation in sweeping magnetic field at a constant coolant temperature of $T\simeq 0.3$ K.

\begin{figure}[t]
\includegraphics{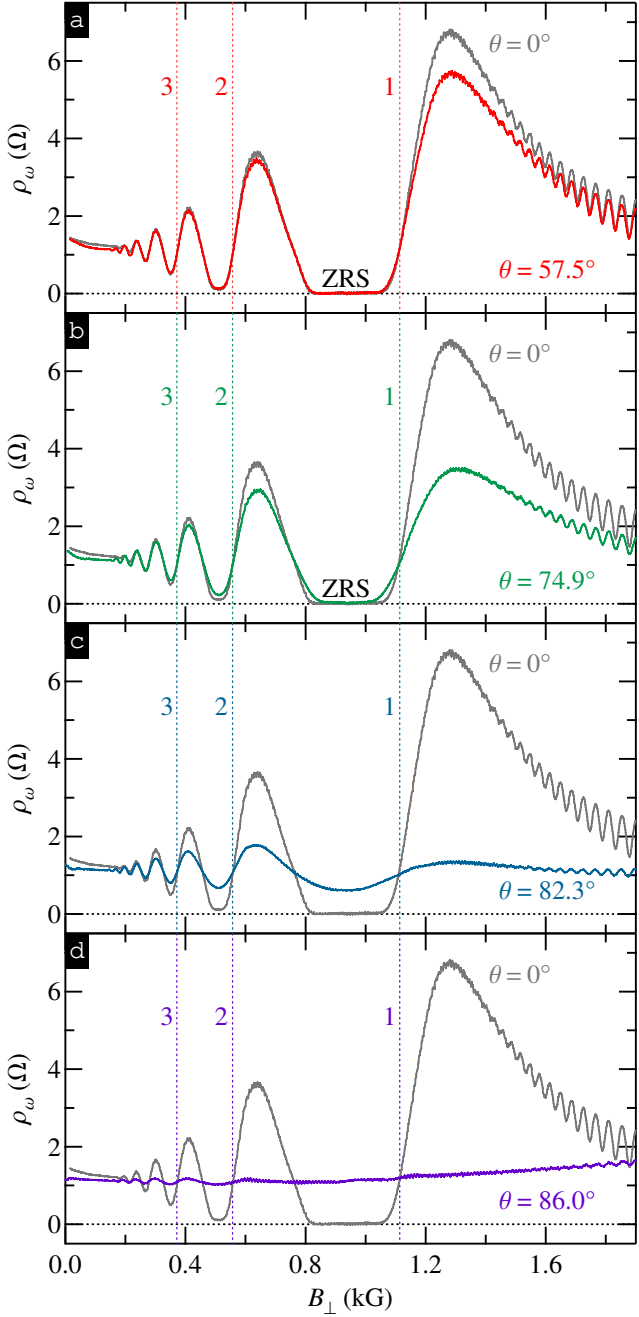}
\caption{(Color online) Resistivity $\rho_\omega(B_\perp)$ measured at $T = 0.3$ K, $f = 48.4$ GHz, and different tilt angles: (a) $\theta = 57.5^\circ$, (b) $\theta = 74.9^\circ$, (c) $\theta = 82.3^\circ$, and (d) $\theta = 86.0^\circ$.
For easy comparison, each panel includes $\rho_\omega(B)$ at $\theta = 0$.
Vertical lines are drawn at the harmonics of the cyclotron resonance.
}
\vspace{-0.15 in}
\label{fig1}
\end{figure}
In \rfig{fig1} we present the magnetoresistivity $\rho_\omega(B_\perp)$ measured under microwave irradiation at different tilt angles: (a) $\theta = 57.5^\circ$, (b) $\theta = 74.9^\circ$, (c) $\theta = 82.3^\circ$, and (d) $\theta = 86.0^\circ$.
The harmonics of the cyclotron resonance are marked by integers (cf.\,$1,\,2,\,3$).
Each panel also includes the data recorded at $\theta = 0^\circ$, which shows pronounced MIRO and a zero-resistance state, attesting to the high quality of our 2DES.

In agreement with previous studies,\citep{mani:2005,yang:2006} we observe that the MIRO period depends only on the perpendicular component of the magnetic field, $B_\perp = B\cos\theta$.
At the same time, direct comparison with the $\theta = 0^\circ$ data reveals that MIRO monotonically decay away with increasing tilt angle. 
As a result, the zero-resistance state is no longer observed at the two highest angles.

The observed decay of MIRO with increasing $\theta$, however, is clearly not uniform and depends sensitively on the oscillation order.
Indeed, in contrast to the data obtained at $\theta = 0^\circ$, where the MIRO amplitude monotonically increases with $B_\perp$, the data obtained at higher angles show more complicated behavior.
For example, the data at $\theta = 57.5^\circ$ and $\theta = 74.9^\circ$ [\rfig{fig1}(a) and \rfig{fig1}(b), respectively], clearly reveal that the first (fundamental) oscillation decays faster than the second, and that there is virtually no change in the strength of higher order oscillations.
At the next angle, $\theta = 82.3^\circ$ [\rfig{fig1}(c)], the first oscillation becomes considerably weaker than the second, while the second oscillation appears roughly the same as the third.
At still higher tilt, $\theta = 86.0^\circ$ [\rfig{fig1}(d)], the lower order oscillations virtually disappear, while the higher order (lower $B_\perp$) oscillations can still be observed.
All these findings indicate that the degree of suppression is determined by an in-plane component of the magnetic field, $B_\parallel$. 
Indeed, since $B_\parallel \propto B_\perp$,\citep{note:1} lower order (higher $B_\perp$) MIRO are subject to a larger $B_\parallel$.

We further notice that titling the sample slightly modifies the background resistance at $B_\perp \lesssim 0.2$ kG.
At higher $B_\perp$, however, there is no noticeable change in the background resistance, as demonstrated by common crossing points of the data with the vertical lines drawn at the cyclotron resonance harmonics.
These points are clearly observed even at the highest tilt angle studied.

\begin{figure}[t]
\includegraphics{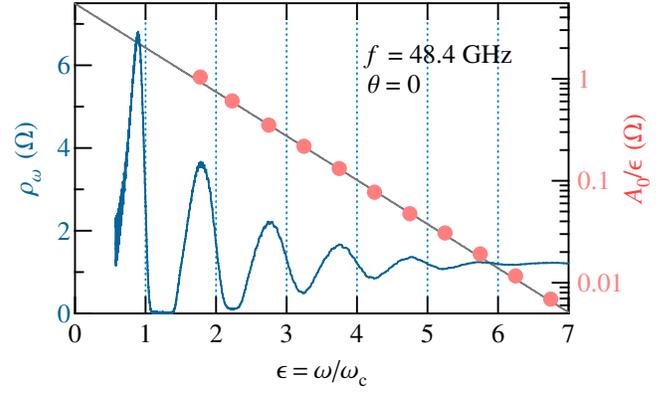}
\caption{(Color online) Resistivity $\rho_\omega$ (left axis) and reduced MIRO amplitude $A_0/\eac$ (right axis) versus $\eac = \omega/\oc$ measured under microwave irradiation of frequency $f=48.4$ GHz at $\theta = 0$.
The fit to the data with $\exp(-\eac/\eac_0)$ (solid line) reveals $\eac_0 = 1.0$ ($\tq^0 \approx 20.7$ ps).
}
\vspace{-0.15 in}
\label{fig2}
\end{figure}
On the left axis of \rfig{fig2} we replot the resistivity $\rho_\omega$ measured at $\theta = 0$ under microwave irradiation as a function of $\eac$.
The oscillations decay {\em monotonically} with increasing $\epsilon$, as prescribed by Eq.\,\ref{eq.miro}.
The smooth part of the $B$ dependence is described by $\eac\lambda^2$, and therefore the reduced MIRO amplitude is described by $A_0/\eac \propto \lambda^2 = \exp(-\eac/\eac_0)$, where $\eac_0 = f\tq^0$.
The experimentally obtained $A_0/\eac$, shown on the right axis of \rfig{fig2}, is very well described by such exponential dependence. 
The fit to the data with $\exp(-\eac/\eac_0)$, shown by the solid line, yields $\eac_0 \approx 1.0$, which corresponds to $\tq^0 \approx 20.7$ ps.

The two parameters which can be affected by $B_\parallel$ are the electron effective mass $m^\star$ and the quantum lifetime $\tq$.
A recent study of Shubnikov-de Haas oscillations (SdHO)in tilted magnetic fields \cite{hatke:2012c} in a similar 2DEG have shown that appreciable change in $m^\star$ calls for $B_\parallel \sim 10^5$ G, which is an order of magnitude higher than $B_\parallel$ used in the present study.
In addition, the change in $m^\star$ would affect the MIRO period, which was not detected even at the highest tilt angle.
We therefore can rule out possible change in $m^\star$ as a source of the observed suppression.

On the other hand, a study of Hall field-induced resistance oscillations \citep{yang:2002,zhang:2007a,hatke:2009c} in tilted magnetic fields \cite{hatke:2011a} has found that the suppression of oscillations can be interpreted in terms of a $B_\parallel$-induced correction to the quantum scattering rate, $\tq^{-1}$.
In particular, it was suggested that in tilted magnetic fields the quantum scattering rate is modified as $1/\tq=1/\tq^0+\delta(1/\tq)$, where $1/\tq^0$ is the scattering rate at $B_\parallel = 0$ and $\delta(1/\tq)$ is the $B_\parallel$-induced correction.
To account for a faster decay at higher $B_\perp$, $\delta(1/\tq)$ should increase faster than $B_\parallel$.\citep{note:2}
Assuming that $\delta(1/\tq) \propto B_\parallel^2$, the $B_\parallel$-induced correction to the argument of the Dingle factor can be written as $-\pi\delta(1/\tq)/\oc\propto -B_\parallel^2/B_\perp \propto -\tan^2\theta/\eac$.
The resultant extra factor to the MIRO amplitude is then given by $\exp(-\alpha\tan^2\theta/\eac)$, where $\alpha$ is a dimensionless constant.
This factor is equal to unity at $\theta = 0$, decreases with $\theta$ (for a given $\eac$) and increases with $\eac$ (for a given $\theta$), consistent with our experimental observations.

\begin{figure}[t]
\includegraphics{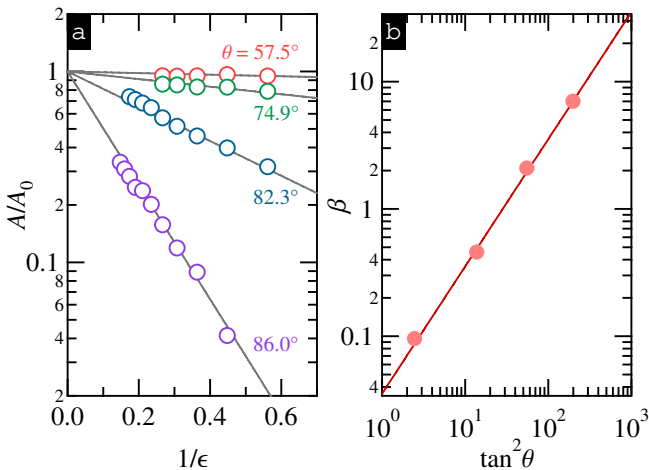}
\caption{(Color online) (a) Normalized MIRO amplitude $A/A_0$ vs $1/\eac$ for $\theta = 57.5^\circ,\,0.260,\,0.133,$ and 0.071 (circles) and fits to $\exp(-\beta\oc/\omega)$ (lines).
(b) Extracted values of $\beta$ vs $\tan^2\theta$ (circles) and a fit, $\beta\simeq 
0.035\cdot\tan^2\theta$ (line).
}
\vspace{-0.15 in}
\label{fig3}
\end{figure}
We next analyze the decay of the MIRO amplitude with increasing $\theta$ in terms of $A(\theta)=A_0\exp(-\beta(\theta)/\eac)$, where $A_0$ is the amplitude at $\theta=0$.
\rFig{fig3}(a) shows the MIRO amplitude, normalized to its value at $\theta = 0$ [cf.\,\rfig{fig2}(b)], $A/A_0$ as a function of $1/\eac$ for different tilt angles, $\theta = 57.5^\circ,\,74.9^\circ,\,82.3^\circ,$ and $86.0^\circ$, as marked.
By fitting these data with $\exp(-\beta/\eac)$ (cf.\,solid lines), we obtain $\beta$ for all tilt angles studied and present the result (circles) in \rfig{fig3}(b) as a function of $\tan^2\theta$ on a log-log scale.
From the linear fit, $\beta =\alpha\tan^2\theta$, (line) we obtain $\alpha \approx 0.035$.
If one writes $\delta(1/\tq) = (1/\tq^0)(B_\parallel/B_0)^2$, obtained $\alpha$ translates to $B_0 \approx 6.0$ kG.
Obtained $B_0$, which corresponds to doubling of the quantum scattering rate in our 2DES, compares well to $B_\parallel \approx 5$ kG which were required to strongly suppress MIRO in \rref{yang:2006}.

\begin{figure}[b]
\includegraphics{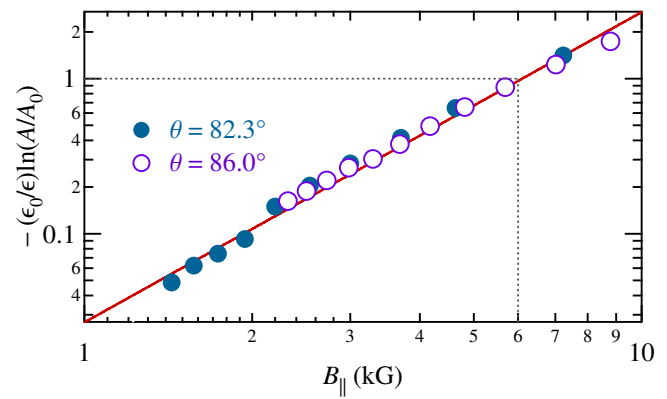}
\caption{(Color online) 
Solid (open) circles are $-(\epsilon_0/\epsilon)\ln(A/A_0)$ for $\theta = 82.3^\circ$ ($\theta = 86.0^\circ$) versus $B_{\parallel}$. 
The solid line corresponds to $(B_\parallel/B_{0})^2$, with $B_{0}= 6.0$ kG.
}
\vspace{-0.15 in}
\label{fig4}
\end{figure}

To further confirm our findings, we notice that, according to \req{eq.miro}, the correction to the zero-tilt scattering rate, $\delta (1/{\tq})$, is directly related to the change of amplitude, as
\begin{equation}
\frac {\delta(1/\tq)}{1/{\tq}^0} = -\frac {\ln(A/A_0)}{\epsilon/\epsilon_0}\,.
\end{equation}
In \rfig{fig4} we present the quantity $-(\epsilon_0/\epsilon)\ln(A/A_0)$, computed from the amplitudes shown in \rfig{fig3}(a), versus $B_{\parallel}$ on a log-log scale. 
Here, solid (open) circles represent amplitudes measured at $\theta = 82.3^\circ$ ($\theta = 86.0^\circ$).
Plotted in such a way, the data for {\em both} tilt angles collapse on a single line extending over nearly two orders of magnitude. 
This line (cf. solid line) is drawn at $(B_\parallel/B_{0})^2$, where $B_{0} = 6.0$ kG. 
We thus conclude that the decay of the MIRO amplitude can be understood in terms of a quadratic-in-$B_\parallel$ change of the quantum scattering rate.

One of the natural questions to ask is whether the obtained enhancement of the quantum scattering rate can also explain the decay of SdHO with increasing $\theta$ (see \rfig{fig1}). 
We first recall that the scattering rate which controls SdHO amplitude is about an order of magnitude larger than that entering the MIRO amplitude.\citep{zudov:2001a,studenikin:2005} 
The difference in the scattering rates is usually attributed to macroscopic density fluctuations, which give rise to extra damping of SdHO (whose period depends on density) but have little effect on MIRO.
It is therefore unlikely that the observed SdHO decay is a result of $\delta(1/\tq)$ discussed above, as it only represents a small fraction of the SdHO scattering rate at $\theta = 0$.
On the other hand, in contrast to MIRO (which are not sensitive to the spin degree of freedom), the SdHO can decay at high tilt angles because of the increased Zeeman energy, which effectively increases the width (and reduces the height) of the initially spin-unresolved Landau levels.\citep{romero:2008}

In summary, we have studied the effect of an in-plane magnetic field on microwave-induced resistance oscillations in a high mobility 2DES.
We have found that the oscillations become progressively weaker as the magnetic field is tilted away from the sample normal.
The rate at which oscillations decay with increasing tilt angle is progressively larger for the lower oscillation orders. 
The analysis shows that the observed decay can be understood in terms of a $B_\parallel$-induced increase of the single particle scattering rate which acquires a quadratic-in-$B_\parallel$ correction, $\delta(1/\tq) = (1/\tq^0)(B_\parallel/B_0)^2$, where $B_0 \approx 6.0$ kG in our 2DES. 
The exact mechanism of such an increase remains a subject of future studies.

The work at Minnesota was supported by the DOE Grant DE-SC0002567.
The work at Princeton was partially funded by the Gordon and Betty Moore Foundation and by the NSF MRSEC Program through the Princeton Center for Complex Materials (DMR-0819860).
A.B. and S.S. acknowledge financial support from NSERC.
A. H. acknowledges support from the NSF Grant No. DMR-0548014.


\begin{thebibliography}{48}
\expandafter\ifx\csname natexlab\endcsname\relax\def\natexlab#1{#1}\fi
\expandafter\ifx\csname bibnamefont\endcsname\relax
  \def\bibnamefont#1{#1}\fi
\expandafter\ifx\csname bibfnamefont\endcsname\relax
  \def\bibfnamefont#1{#1}\fi
\expandafter\ifx\csname citenamefont\endcsname\relax
  \def\citenamefont#1{#1}\fi
\expandafter\ifx\csname url\endcsname\relax
  \def\url#1{\texttt{#1}}\fi
\expandafter\ifx\csname urlprefix\endcsname\relax\def\urlprefix{URL }\fi
\providecommand{\bibinfo}[2]{#2}
\providecommand{\eprint}[2][]{\url{#2}}

\bibitem[{\citenamefont{Zudov et~al.}(2001{\natexlab{a}})\citenamefont{Zudov,
  Du, Simmons, and Reno}}]{zudov:2001a}
\bibinfo{author}{\bibfnamefont{M.~A.} \bibnamefont{Zudov}},
  \bibinfo{author}{\bibfnamefont{R.~R.} \bibnamefont{Du}},
  \bibinfo{author}{\bibfnamefont{J.~A.} \bibnamefont{Simmons}},
  \bibnamefont{and} \bibinfo{author}{\bibfnamefont{J.~L.} \bibnamefont{Reno}},
  \bibinfo{journal}{Phys. Rev. B} \textbf{\bibinfo{volume}{64}},
  \bibinfo{pages}{201311(R)} (\bibinfo{year}{2001}{\natexlab{a}}).

\bibitem[{\citenamefont{Ye et~al.}(2001)\citenamefont{Ye, Engel, Tsui, Simmons,
  Wendt et~al.}}]{ye:2001}
\bibinfo{author}{\bibfnamefont{P.~D.} \bibnamefont{Ye}},
  \bibinfo{author}{\bibfnamefont{L.~W.} \bibnamefont{Engel}},
  \bibinfo{author}{\bibfnamefont{D.~C.} \bibnamefont{Tsui}},
  \bibinfo{author}{\bibfnamefont{J.~A.} \bibnamefont{Simmons}},
  \bibinfo{author}{\bibfnamefont{J.~R.} \bibnamefont{Wendt}},
  \bibnamefont{et~al.}, \bibinfo{journal}{Appl. Phys. Lett.}
  \textbf{\bibinfo{volume}{79}}, \bibinfo{pages}{2193} (\bibinfo{year}{2001}).

\bibitem[{\citenamefont{Mani et~al.}(2002)\citenamefont{Mani, Smet, von
  Klitzing, Narayanamurti, Johnson et~al.}}]{mani:2002}
\bibinfo{author}{\bibfnamefont{R.~G.} \bibnamefont{Mani}},
  \bibinfo{author}{\bibfnamefont{J.~H.} \bibnamefont{Smet}},
  \bibinfo{author}{\bibfnamefont{K.}~\bibnamefont{von Klitzing}},
  \bibinfo{author}{\bibfnamefont{V.}~\bibnamefont{Narayanamurti}},
  \bibinfo{author}{\bibfnamefont{W.~B.} \bibnamefont{Johnson}},
  \bibnamefont{et~al.}, \bibinfo{journal}{Nature (London)}
  \textbf{\bibinfo{volume}{420}}, \bibinfo{pages}{646} (\bibinfo{year}{2002}).

\bibitem[{\citenamefont{Zudov et~al.}(2003)\citenamefont{Zudov, Du, Pfeiffer,
  and West}}]{zudov:2003}
\bibinfo{author}{\bibfnamefont{M.~A.} \bibnamefont{Zudov}},
  \bibinfo{author}{\bibfnamefont{R.~R.} \bibnamefont{Du}},
  \bibinfo{author}{\bibfnamefont{L.~N.} \bibnamefont{Pfeiffer}},
  \bibnamefont{and} \bibinfo{author}{\bibfnamefont{K.~W.} \bibnamefont{West}},
  \bibinfo{journal}{Phys. Rev. Lett.} \textbf{\bibinfo{volume}{90}},
  \bibinfo{pages}{046807} (\bibinfo{year}{2003}).

\bibitem[{\citenamefont{Zudov et~al.}(2001{\natexlab{b}})\citenamefont{Zudov,
  Ponomarev, Efros, Du, Simmons et~al.}}]{zudov:2001b}
\bibinfo{author}{\bibfnamefont{M.~A.} \bibnamefont{Zudov}},
  \bibinfo{author}{\bibfnamefont{I.~V.} \bibnamefont{Ponomarev}},
  \bibinfo{author}{\bibfnamefont{A.~L.} \bibnamefont{Efros}},
  \bibinfo{author}{\bibfnamefont{R.~R.} \bibnamefont{Du}},
  \bibinfo{author}{\bibfnamefont{J.~A.} \bibnamefont{Simmons}},
  \bibnamefont{et~al.}, \bibinfo{journal}{Phys. Rev. Lett.}
  \textbf{\bibinfo{volume}{86}}, \bibinfo{pages}{3614}
  (\bibinfo{year}{2001}{\natexlab{b}}).

\bibitem[{\citenamefont{Yang et~al.}(2002)\citenamefont{Yang, Zhang, Du,
  Simmons, and Reno}}]{yang:2002}
\bibinfo{author}{\bibfnamefont{C.~L.} \bibnamefont{Yang}},
  \bibinfo{author}{\bibfnamefont{J.}~\bibnamefont{Zhang}},
  \bibinfo{author}{\bibfnamefont{R.~R.} \bibnamefont{Du}},
  \bibinfo{author}{\bibfnamefont{J.~A.} \bibnamefont{Simmons}},
  \bibnamefont{and} \bibinfo{author}{\bibfnamefont{J.~L.} \bibnamefont{Reno}},
  \bibinfo{journal}{Phys. Rev. Lett.} \textbf{\bibinfo{volume}{89}},
  \bibinfo{pages}{076801} (\bibinfo{year}{2002}).

\bibitem[{\citenamefont{Zhang et~al.}(2007{\natexlab{a}})\citenamefont{Zhang,
  Zudov, Pfeiffer, and West}}]{zhang:2007c}
\bibinfo{author}{\bibfnamefont{W.}~\bibnamefont{Zhang}},
  \bibinfo{author}{\bibfnamefont{M.~A.} \bibnamefont{Zudov}},
  \bibinfo{author}{\bibfnamefont{L.~N.} \bibnamefont{Pfeiffer}},
  \bibnamefont{and} \bibinfo{author}{\bibfnamefont{K.~W.} \bibnamefont{West}},
  \bibinfo{journal}{Phys. Rev. Lett.} \textbf{\bibinfo{volume}{98}},
  \bibinfo{pages}{106804} (\bibinfo{year}{2007}{\natexlab{a}}).

\bibitem[{\citenamefont{Zhang et~al.}(2008)\citenamefont{Zhang, Zudov,
  Pfeiffer, and West}}]{zhang:2008}
\bibinfo{author}{\bibfnamefont{W.}~\bibnamefont{Zhang}},
  \bibinfo{author}{\bibfnamefont{M.~A.} \bibnamefont{Zudov}},
  \bibinfo{author}{\bibfnamefont{L.~N.} \bibnamefont{Pfeiffer}},
  \bibnamefont{and} \bibinfo{author}{\bibfnamefont{K.~W.} \bibnamefont{West}},
  \bibinfo{journal}{Phys. Rev. Lett.} \textbf{\bibinfo{volume}{100}},
  \bibinfo{pages}{036805} (\bibinfo{year}{2008}).

\bibitem[{\citenamefont{Khodas et~al.}(2010)\citenamefont{Khodas, Chiang,
  Hatke, Zudov, Vavilov et~al.}}]{khodas:2010}
\bibinfo{author}{\bibfnamefont{M.}~\bibnamefont{Khodas}},
  \bibinfo{author}{\bibfnamefont{H.~S.} \bibnamefont{Chiang}},
  \bibinfo{author}{\bibfnamefont{A.~T.} \bibnamefont{Hatke}},
  \bibinfo{author}{\bibfnamefont{M.~A.} \bibnamefont{Zudov}},
  \bibinfo{author}{\bibfnamefont{M.~G.} \bibnamefont{Vavilov}},
  \bibnamefont{et~al.}, \bibinfo{journal}{Phys. Rev. Lett.}
  \textbf{\bibinfo{volume}{104}}, \bibinfo{pages}{206801}
  (\bibinfo{year}{2010}).

\bibitem[{\citenamefont{Wiedmann et~al.}(2010)\citenamefont{Wiedmann, Gusev,
  Raichev, Bakarov, and Portal}}]{wiedmann:2010b}
\bibinfo{author}{\bibfnamefont{S.}~\bibnamefont{Wiedmann}},
  \bibinfo{author}{\bibfnamefont{G.~M.} \bibnamefont{Gusev}},
  \bibinfo{author}{\bibfnamefont{O.~E.} \bibnamefont{Raichev}},
  \bibinfo{author}{\bibfnamefont{A.~K.} \bibnamefont{Bakarov}},
  \bibnamefont{and} \bibinfo{author}{\bibfnamefont{J.~C.}
  \bibnamefont{Portal}}, \bibinfo{journal}{Phys. Rev. Lett.}
  \textbf{\bibinfo{volume}{105}}, \bibinfo{pages}{026804}
  (\bibinfo{year}{2010}).

\bibitem[{\citenamefont{Hatke et~al.}(2010)\citenamefont{Hatke, Chiang, Zudov,
  Pfeiffer, and West}}]{hatke:2010a}
\bibinfo{author}{\bibfnamefont{A.~T.} \bibnamefont{Hatke}},
  \bibinfo{author}{\bibfnamefont{H.-S.} \bibnamefont{Chiang}},
  \bibinfo{author}{\bibfnamefont{M.~A.} \bibnamefont{Zudov}},
  \bibinfo{author}{\bibfnamefont{L.~N.} \bibnamefont{Pfeiffer}},
  \bibnamefont{and} \bibinfo{author}{\bibfnamefont{K.~W.} \bibnamefont{West}},
  \bibinfo{journal}{Phys. Rev. B} \textbf{\bibinfo{volume}{82}},
  \bibinfo{pages}{041304(R)} (\bibinfo{year}{2010}).

\bibitem[{\citenamefont{Dai et~al.}(2010)\citenamefont{Dai, Du, Pfeiffer, and
  West}}]{dai:2010}
\bibinfo{author}{\bibfnamefont{Y.}~\bibnamefont{Dai}},
  \bibinfo{author}{\bibfnamefont{R.~R.} \bibnamefont{Du}},
  \bibinfo{author}{\bibfnamefont{L.~N.} \bibnamefont{Pfeiffer}},
  \bibnamefont{and} \bibinfo{author}{\bibfnamefont{K.~W.} \bibnamefont{West}},
  \bibinfo{journal}{Phys. Rev. Lett.} \textbf{\bibinfo{volume}{105}},
  \bibinfo{pages}{246802} (\bibinfo{year}{2010}).

\bibitem[{\citenamefont{Hatke et~al.}(2011{\natexlab{a}})\citenamefont{Hatke,
  Zudov, Pfeiffer, and West}}]{hatke:2011b}
\bibinfo{author}{\bibfnamefont{A.~T.} \bibnamefont{Hatke}},
  \bibinfo{author}{\bibfnamefont{M.~A.} \bibnamefont{Zudov}},
  \bibinfo{author}{\bibfnamefont{L.~N.} \bibnamefont{Pfeiffer}},
  \bibnamefont{and} \bibinfo{author}{\bibfnamefont{K.~W.} \bibnamefont{West}},
  \bibinfo{journal}{Phys. Rev. B} \textbf{\bibinfo{volume}{83}},
  \bibinfo{pages}{121301(R)} (\bibinfo{year}{2011}{\natexlab{a}}).

\bibitem[{\citenamefont{Andreev et~al.}(2003)\citenamefont{Andreev, Aleiner,
  and Millis}}]{andreev:2003}
\bibinfo{author}{\bibfnamefont{A.~V.} \bibnamefont{Andreev}},
  \bibinfo{author}{\bibfnamefont{I.~L.} \bibnamefont{Aleiner}},
  \bibnamefont{and} \bibinfo{author}{\bibfnamefont{A.~J.}
  \bibnamefont{Millis}}, \bibinfo{journal}{Phys. Rev. Lett.}
  \textbf{\bibinfo{volume}{91}}, \bibinfo{pages}{056803}
  (\bibinfo{year}{2003}).

\bibitem[{\citenamefont{Durst et~al.}(2003)\citenamefont{Durst, Sachdev, Read,
  and Girvin}}]{durst:2003}
\bibinfo{author}{\bibfnamefont{A.~C.} \bibnamefont{Durst}},
  \bibinfo{author}{\bibfnamefont{S.}~\bibnamefont{Sachdev}},
  \bibinfo{author}{\bibfnamefont{N.}~\bibnamefont{Read}}, \bibnamefont{and}
  \bibinfo{author}{\bibfnamefont{S.~M.} \bibnamefont{Girvin}},
  \bibinfo{journal}{Phys. Rev. Lett.} \textbf{\bibinfo{volume}{91}},
  \bibinfo{pages}{086803} (\bibinfo{year}{2003}).

\bibitem[{\citenamefont{Lei and Liu}(2003)}]{lei:2003}
\bibinfo{author}{\bibfnamefont{X.~L.} \bibnamefont{Lei}} \bibnamefont{and}
  \bibinfo{author}{\bibfnamefont{S.~Y.} \bibnamefont{Liu}},
  \bibinfo{journal}{Phys. Rev. Lett.} \textbf{\bibinfo{volume}{91}},
  \bibinfo{pages}{226805} (\bibinfo{year}{2003}).

\bibitem[{\citenamefont{Vavilov and Aleiner}(2004)}]{vavilov:2004}
\bibinfo{author}{\bibfnamefont{M.~G.} \bibnamefont{Vavilov}} \bibnamefont{and}
  \bibinfo{author}{\bibfnamefont{I.~L.} \bibnamefont{Aleiner}},
  \bibinfo{journal}{Phys. Rev. B} \textbf{\bibinfo{volume}{69}},
  \bibinfo{pages}{035303} (\bibinfo{year}{2004}).

\bibitem[{\citenamefont{Dmitriev et~al.}(2003)\citenamefont{Dmitriev, Mirlin,
  and Polyakov}}]{dmitriev:2003}
\bibinfo{author}{\bibfnamefont{I.~A.} \bibnamefont{Dmitriev}},
  \bibinfo{author}{\bibfnamefont{A.~D.} \bibnamefont{Mirlin}},
  \bibnamefont{and} \bibinfo{author}{\bibfnamefont{D.~G.}
  \bibnamefont{Polyakov}}, \bibinfo{journal}{Phys. Rev. Lett.}
  \textbf{\bibinfo{volume}{91}}, \bibinfo{pages}{226802}
  (\bibinfo{year}{2003}).

\bibitem[{\citenamefont{Dmitriev et~al.}(2005)\citenamefont{Dmitriev, Vavilov,
  Aleiner, Mirlin, and Polyakov}}]{dmitriev:2005}
\bibinfo{author}{\bibfnamefont{I.~A.} \bibnamefont{Dmitriev}},
  \bibinfo{author}{\bibfnamefont{M.~G.} \bibnamefont{Vavilov}},
  \bibinfo{author}{\bibfnamefont{I.~L.} \bibnamefont{Aleiner}},
  \bibinfo{author}{\bibfnamefont{A.~D.} \bibnamefont{Mirlin}},
  \bibnamefont{and} \bibinfo{author}{\bibfnamefont{D.~G.}
  \bibnamefont{Polyakov}}, \bibinfo{journal}{Phys. Rev. B}
  \textbf{\bibinfo{volume}{71}}, \bibinfo{pages}{115316}
  (\bibinfo{year}{2005}).

\bibitem[{\citenamefont{Khodas and Vavilov}(2008)}]{khodas:2008}
\bibinfo{author}{\bibfnamefont{M.}~\bibnamefont{Khodas}} \bibnamefont{and}
  \bibinfo{author}{\bibfnamefont{M.~G.} \bibnamefont{Vavilov}},
  \bibinfo{journal}{Phys. Rev. B} \textbf{\bibinfo{volume}{78}},
  \bibinfo{pages}{245319} (\bibinfo{year}{2008}).

\bibitem[{\citenamefont{Dmitriev et~al.}(2009)\citenamefont{Dmitriev, Khodas,
  Mirlin, Polyakov, and Vavilov}}]{dmitriev:2009b}
\bibinfo{author}{\bibfnamefont{I.~A.} \bibnamefont{Dmitriev}},
  \bibinfo{author}{\bibfnamefont{M.}~\bibnamefont{Khodas}},
  \bibinfo{author}{\bibfnamefont{A.~D.} \bibnamefont{Mirlin}},
  \bibinfo{author}{\bibfnamefont{D.~G.} \bibnamefont{Polyakov}},
  \bibnamefont{and} \bibinfo{author}{\bibfnamefont{M.~G.}
  \bibnamefont{Vavilov}}, \bibinfo{journal}{Phys. Rev. B}
  \textbf{\bibinfo{volume}{80}}, \bibinfo{pages}{165327}
  (\bibinfo{year}{2009}).

\bibitem[{\citenamefont{Hatke et~al.}(2011{\natexlab{b}})\citenamefont{Hatke,
  Khodas, Zudov, Pfeiffer, and West}}]{hatke:2011e}
\bibinfo{author}{\bibfnamefont{A.~T.} \bibnamefont{Hatke}},
  \bibinfo{author}{\bibfnamefont{M.}~\bibnamefont{Khodas}},
  \bibinfo{author}{\bibfnamefont{M.~A.} \bibnamefont{Zudov}},
  \bibinfo{author}{\bibfnamefont{L.~N.} \bibnamefont{Pfeiffer}},
  \bibnamefont{and} \bibinfo{author}{\bibfnamefont{K.~W.} \bibnamefont{West}},
  \bibinfo{journal}{Phys. Rev. B} \textbf{\bibinfo{volume}{84}},
  \bibinfo{pages}{241302(R)} (\bibinfo{year}{2011}{\natexlab{b}}).

\bibitem[{\citenamefont{Zudov}(2004)}]{zudov:2004}
\bibinfo{author}{\bibfnamefont{M.~A.} \bibnamefont{Zudov}},
  \bibinfo{journal}{Phys. Rev. B} \textbf{\bibinfo{volume}{69}},
  \bibinfo{pages}{041304(R)} (\bibinfo{year}{2004}).

\bibitem[{\citenamefont{Willett et~al.}(2004)\citenamefont{Willett, Pfeiffer,
  and West}}]{willett:2004}
\bibinfo{author}{\bibfnamefont{R.~L.} \bibnamefont{Willett}},
  \bibinfo{author}{\bibfnamefont{L.~N.} \bibnamefont{Pfeiffer}},
  \bibnamefont{and} \bibinfo{author}{\bibfnamefont{K.~W.} \bibnamefont{West}},
  \bibinfo{journal}{Phys. Rev. Lett.} \textbf{\bibinfo{volume}{93}},
  \bibinfo{pages}{026804} (\bibinfo{year}{2004}).

\bibitem[{\citenamefont{Mani et~al.}(2004{\natexlab{a}})\citenamefont{Mani,
  Narayanamurti, von Klitzing, Smet, Johnson et~al.}}]{mani:2004a}
\bibinfo{author}{\bibfnamefont{R.~G.} \bibnamefont{Mani}},
  \bibinfo{author}{\bibfnamefont{V.}~\bibnamefont{Narayanamurti}},
  \bibinfo{author}{\bibfnamefont{K.}~\bibnamefont{von Klitzing}},
  \bibinfo{author}{\bibfnamefont{J.~H.} \bibnamefont{Smet}},
  \bibinfo{author}{\bibfnamefont{W.~B.} \bibnamefont{Johnson}},
  \bibnamefont{et~al.}, \bibinfo{journal}{Phys. Rev. B}
  \textbf{\bibinfo{volume}{70}}, \bibinfo{pages}{155310}
  (\bibinfo{year}{2004}{\natexlab{a}}).

\bibitem[{\citenamefont{Mani et~al.}(2004{\natexlab{b}})\citenamefont{Mani,
  Smet, von Klitzing, Narayanamurti, Johnson et~al.}}]{mani:2004e}
\bibinfo{author}{\bibfnamefont{R.~G.} \bibnamefont{Mani}},
  \bibinfo{author}{\bibfnamefont{J.~H.} \bibnamefont{Smet}},
  \bibinfo{author}{\bibfnamefont{K.}~\bibnamefont{von Klitzing}},
  \bibinfo{author}{\bibfnamefont{V.}~\bibnamefont{Narayanamurti}},
  \bibinfo{author}{\bibfnamefont{W.~B.} \bibnamefont{Johnson}},
  \bibnamefont{et~al.}, \bibinfo{journal}{Phys. Rev. Lett.}
  \textbf{\bibinfo{volume}{92}}, \bibinfo{pages}{146801}
  (\bibinfo{year}{2004}{\natexlab{b}}).

\bibitem[{\citenamefont{Studenikin et~al.}(2004)\citenamefont{Studenikin,
  Potemski, Coleridge, Sachrajda, and Wasilewski}}]{studenikin:2004}
\bibinfo{author}{\bibfnamefont{S.~A.} \bibnamefont{Studenikin}},
  \bibinfo{author}{\bibfnamefont{M.}~\bibnamefont{Potemski}},
  \bibinfo{author}{\bibfnamefont{P.~T.} \bibnamefont{Coleridge}},
  \bibinfo{author}{\bibfnamefont{A.~S.} \bibnamefont{Sachrajda}},
  \bibnamefont{and} \bibinfo{author}{\bibfnamefont{Z.~R.}
  \bibnamefont{Wasilewski}}, \bibinfo{journal}{Solid State Commun.}
  \textbf{\bibinfo{volume}{129}}, \bibinfo{pages}{341} (\bibinfo{year}{2004}).

\bibitem[{\citenamefont{Studenikin et~al.}(2005)\citenamefont{Studenikin,
  Potemski, Sachrajda, Hilke, Pfeiffer et~al.}}]{studenikin:2005}
\bibinfo{author}{\bibfnamefont{S.~A.} \bibnamefont{Studenikin}},
  \bibinfo{author}{\bibfnamefont{M.}~\bibnamefont{Potemski}},
  \bibinfo{author}{\bibfnamefont{A.}~\bibnamefont{Sachrajda}},
  \bibinfo{author}{\bibfnamefont{M.}~\bibnamefont{Hilke}},
  \bibinfo{author}{\bibfnamefont{L.~N.} \bibnamefont{Pfeiffer}},
  \bibnamefont{et~al.}, \bibinfo{journal}{Phys. Rev. B}
  \textbf{\bibinfo{volume}{71}}, \bibinfo{pages}{245313}
  (\bibinfo{year}{2005}).

\bibitem[{\citenamefont{Smet et~al.}(2005)\citenamefont{Smet, Gorshunov, Jiang,
  Pfeiffer, West et~al.}}]{smet:2005}
\bibinfo{author}{\bibfnamefont{J.~H.} \bibnamefont{Smet}},
  \bibinfo{author}{\bibfnamefont{B.}~\bibnamefont{Gorshunov}},
  \bibinfo{author}{\bibfnamefont{C.}~\bibnamefont{Jiang}},
  \bibinfo{author}{\bibfnamefont{L.}~\bibnamefont{Pfeiffer}},
  \bibinfo{author}{\bibfnamefont{K.}~\bibnamefont{West}}, \bibnamefont{et~al.},
  \bibinfo{journal}{Phys. Rev. Lett.} \textbf{\bibinfo{volume}{95}},
  \bibinfo{pages}{116804} (\bibinfo{year}{2005}).

\bibitem[{\citenamefont{Mani}(2005)}]{mani:2005}
\bibinfo{author}{\bibfnamefont{R.~G.} \bibnamefont{Mani}},
  \bibinfo{journal}{Phys. Rev. B} \textbf{\bibinfo{volume}{72}},
  \bibinfo{pages}{075327} (\bibinfo{year}{2005}).

\bibitem[{\citenamefont{Yang et~al.}(2006)\citenamefont{Yang, Du, Pfeiffer, and
  West}}]{yang:2006}
\bibinfo{author}{\bibfnamefont{C.~L.} \bibnamefont{Yang}},
  \bibinfo{author}{\bibfnamefont{R.~R.} \bibnamefont{Du}},
  \bibinfo{author}{\bibfnamefont{L.~N.} \bibnamefont{Pfeiffer}},
  \bibnamefont{and} \bibinfo{author}{\bibfnamefont{K.~W.} \bibnamefont{West}},
  \bibinfo{journal}{Phys. Rev. B} \textbf{\bibinfo{volume}{74}},
  \bibinfo{pages}{045315} (\bibinfo{year}{2006}).

\bibitem[{\citenamefont{Zudov et~al.}(2006{\natexlab{a}})\citenamefont{Zudov,
  Du, Pfeiffer, and West}}]{zudov:2006a}
\bibinfo{author}{\bibfnamefont{M.~A.} \bibnamefont{Zudov}},
  \bibinfo{author}{\bibfnamefont{R.~R.} \bibnamefont{Du}},
  \bibinfo{author}{\bibfnamefont{L.~N.} \bibnamefont{Pfeiffer}},
  \bibnamefont{and} \bibinfo{author}{\bibfnamefont{K.~W.} \bibnamefont{West}},
  \bibinfo{journal}{Phys. Rev. B} \textbf{\bibinfo{volume}{73}},
  \bibinfo{pages}{041303(R)} (\bibinfo{year}{2006}{\natexlab{a}}).

\bibitem[{\citenamefont{Zudov et~al.}(2006{\natexlab{b}})\citenamefont{Zudov,
  Du, Pfeiffer, and West}}]{zudov:2006b}
\bibinfo{author}{\bibfnamefont{M.~A.} \bibnamefont{Zudov}},
  \bibinfo{author}{\bibfnamefont{R.~R.} \bibnamefont{Du}},
  \bibinfo{author}{\bibfnamefont{L.~N.} \bibnamefont{Pfeiffer}},
  \bibnamefont{and} \bibinfo{author}{\bibfnamefont{K.~W.} \bibnamefont{West}},
  \bibinfo{journal}{Phys. Rev. Lett.} \textbf{\bibinfo{volume}{96}},
  \bibinfo{pages}{236804} (\bibinfo{year}{2006}{\natexlab{b}}).

\bibitem[{\citenamefont{Studenikin et~al.}(2007)\citenamefont{Studenikin,
  Sachrajda, Gupta, Wasilewski, Fedorych et~al.}}]{studenikin:2007}
\bibinfo{author}{\bibfnamefont{S.~A.} \bibnamefont{Studenikin}},
  \bibinfo{author}{\bibfnamefont{A.~S.} \bibnamefont{Sachrajda}},
  \bibinfo{author}{\bibfnamefont{J.~A.} \bibnamefont{Gupta}},
  \bibinfo{author}{\bibfnamefont{Z.~R.} \bibnamefont{Wasilewski}},
  \bibinfo{author}{\bibfnamefont{O.~M.} \bibnamefont{Fedorych}},
  \bibnamefont{et~al.}, \bibinfo{journal}{Phys. Rev. B}
  \textbf{\bibinfo{volume}{76}}, \bibinfo{pages}{165321}
  (\bibinfo{year}{2007}).

\bibitem[{\citenamefont{Hatke et~al.}(2008{\natexlab{a}})\citenamefont{Hatke,
  Chiang, Zudov, Pfeiffer, and West}}]{hatke:2008a}
\bibinfo{author}{\bibfnamefont{A.~T.} \bibnamefont{Hatke}},
  \bibinfo{author}{\bibfnamefont{H.-S.} \bibnamefont{Chiang}},
  \bibinfo{author}{\bibfnamefont{M.~A.} \bibnamefont{Zudov}},
  \bibinfo{author}{\bibfnamefont{L.~N.} \bibnamefont{Pfeiffer}},
  \bibnamefont{and} \bibinfo{author}{\bibfnamefont{K.~W.} \bibnamefont{West}},
  \bibinfo{journal}{Phys. Rev. B} \textbf{\bibinfo{volume}{77}},
  \bibinfo{pages}{201304(R)} (\bibinfo{year}{2008}{\natexlab{a}}).

\bibitem[{\citenamefont{Hatke et~al.}(2008{\natexlab{b}})\citenamefont{Hatke,
  Chiang, Zudov, Pfeiffer, and West}}]{hatke:2008b}
\bibinfo{author}{\bibfnamefont{A.~T.} \bibnamefont{Hatke}},
  \bibinfo{author}{\bibfnamefont{H.-S.} \bibnamefont{Chiang}},
  \bibinfo{author}{\bibfnamefont{M.~A.} \bibnamefont{Zudov}},
  \bibinfo{author}{\bibfnamefont{L.~N.} \bibnamefont{Pfeiffer}},
  \bibnamefont{and} \bibinfo{author}{\bibfnamefont{K.~W.} \bibnamefont{West}},
  \bibinfo{journal}{Phys. Rev. Lett.} \textbf{\bibinfo{volume}{101}},
  \bibinfo{pages}{246811} (\bibinfo{year}{2008}{\natexlab{b}}).

\bibitem[{\citenamefont{Hatke et~al.}(2009{\natexlab{a}})\citenamefont{Hatke,
  Zudov, Pfeiffer, and West}}]{hatke:2009a}
\bibinfo{author}{\bibfnamefont{A.~T.} \bibnamefont{Hatke}},
  \bibinfo{author}{\bibfnamefont{M.~A.} \bibnamefont{Zudov}},
  \bibinfo{author}{\bibfnamefont{L.~N.} \bibnamefont{Pfeiffer}},
  \bibnamefont{and} \bibinfo{author}{\bibfnamefont{K.~W.} \bibnamefont{West}},
  \bibinfo{journal}{Phys. Rev. Lett.} \textbf{\bibinfo{volume}{102}},
  \bibinfo{pages}{066804} (\bibinfo{year}{2009}{\natexlab{a}}).

\bibitem[{\citenamefont{Mani et~al.}(2010)\citenamefont{Mani, Gerl, Schmult,
  Wegscheider, and Umansky}}]{mani:2010}
\bibinfo{author}{\bibfnamefont{R.~G.} \bibnamefont{Mani}},
  \bibinfo{author}{\bibfnamefont{C.}~\bibnamefont{Gerl}},
  \bibinfo{author}{\bibfnamefont{S.}~\bibnamefont{Schmult}},
  \bibinfo{author}{\bibfnamefont{W.}~\bibnamefont{Wegscheider}},
  \bibnamefont{and} \bibinfo{author}{\bibfnamefont{V.}~\bibnamefont{Umansky}},
  \bibinfo{journal}{Phys. Rev. B} \textbf{\bibinfo{volume}{81}},
  \bibinfo{pages}{125320} (\bibinfo{year}{2010}).

\bibitem[{\citenamefont{Dorozhkin et~al.}(2011)\citenamefont{Dorozhkin,
  Pfeiffer, West, von Klitzing, and Smet}}]{dorozhkin:2011}
\bibinfo{author}{\bibfnamefont{S.~I.} \bibnamefont{Dorozhkin}},
  \bibinfo{author}{\bibfnamefont{L.}~\bibnamefont{Pfeiffer}},
  \bibinfo{author}{\bibfnamefont{K.}~\bibnamefont{West}},
  \bibinfo{author}{\bibfnamefont{K.}~\bibnamefont{von Klitzing}},
  \bibnamefont{and} \bibinfo{author}{\bibfnamefont{J.~H.} \bibnamefont{Smet}},
  \bibinfo{journal}{Nature Phys.} \textbf{\bibinfo{volume}{7}},
  \bibinfo{pages}{336} (\bibinfo{year}{2011}).

\bibitem[{\citenamefont{Hatke et~al.}(2011{\natexlab{c}})\citenamefont{Hatke,
  Zudov, Pfeiffer, and West}}]{hatke:2011c}
\bibinfo{author}{\bibfnamefont{A.~T.} \bibnamefont{Hatke}},
  \bibinfo{author}{\bibfnamefont{M.~A.} \bibnamefont{Zudov}},
  \bibinfo{author}{\bibfnamefont{L.~N.} \bibnamefont{Pfeiffer}},
  \bibnamefont{and} \bibinfo{author}{\bibfnamefont{K.~W.} \bibnamefont{West}},
  \bibinfo{journal}{Phys. Rev. B} \textbf{\bibinfo{volume}{83}},
  \bibinfo{pages}{201301(R)} (\bibinfo{year}{2011}{\natexlab{c}}).

\bibitem[{\citenamefont{Hatke et~al.}(2011{\natexlab{d}})\citenamefont{Hatke,
  Zudov, Pfeiffer, and West}}]{hatke:2011f}
\bibinfo{author}{\bibfnamefont{A.~T.} \bibnamefont{Hatke}},
  \bibinfo{author}{\bibfnamefont{M.~A.} \bibnamefont{Zudov}},
  \bibinfo{author}{\bibfnamefont{L.~N.} \bibnamefont{Pfeiffer}},
  \bibnamefont{and} \bibinfo{author}{\bibfnamefont{K.~W.} \bibnamefont{West}},
  \bibinfo{journal}{Phys. Rev. B} \textbf{\bibinfo{volume}{84}},
  \bibinfo{pages}{241304(R)} (\bibinfo{year}{2011}{\natexlab{d}}).

\bibitem[{not({\natexlab{a}})}]{note:1}
\bibinfo{note}{For a given $\theta$, $B_\parallel = B \sin \theta =
  B_\perp/\tan\theta$.}

\bibitem[{\citenamefont{Hatke et~al.}(2012)\citenamefont{Hatke, Zudov,
  Pfeiffer, and West}}]{hatke:2012c}
\bibinfo{author}{\bibfnamefont{A.~T.} \bibnamefont{Hatke}},
  \bibinfo{author}{\bibfnamefont{M.~A.} \bibnamefont{Zudov}},
  \bibinfo{author}{\bibfnamefont{L.~N.} \bibnamefont{Pfeiffer}},
  \bibnamefont{and} \bibinfo{author}{\bibfnamefont{K.~W.} \bibnamefont{West}},
  \bibinfo{journal}{Phys. Rev. B} \textbf{\bibinfo{volume}{85}},
  \bibinfo{pages}{241305(R)} (\bibinfo{year}{2012}).

\bibitem[{\citenamefont{Zhang et~al.}(2007{\natexlab{b}})\citenamefont{Zhang,
  Chiang, Zudov, Pfeiffer, and West}}]{zhang:2007a}
\bibinfo{author}{\bibfnamefont{W.}~\bibnamefont{Zhang}},
  \bibinfo{author}{\bibfnamefont{H.-S.} \bibnamefont{Chiang}},
  \bibinfo{author}{\bibfnamefont{M.~A.} \bibnamefont{Zudov}},
  \bibinfo{author}{\bibfnamefont{L.~N.} \bibnamefont{Pfeiffer}},
  \bibnamefont{and} \bibinfo{author}{\bibfnamefont{K.~W.} \bibnamefont{West}},
  \bibinfo{journal}{Phys. Rev. B} \textbf{\bibinfo{volume}{75}},
  \bibinfo{pages}{041304(R)} (\bibinfo{year}{2007}{\natexlab{b}}).

\bibitem[{\citenamefont{Hatke et~al.}(2009{\natexlab{b}})\citenamefont{Hatke,
  Zudov, Pfeiffer, and West}}]{hatke:2009c}
\bibinfo{author}{\bibfnamefont{A.~T.} \bibnamefont{Hatke}},
  \bibinfo{author}{\bibfnamefont{M.~A.} \bibnamefont{Zudov}},
  \bibinfo{author}{\bibfnamefont{L.~N.} \bibnamefont{Pfeiffer}},
  \bibnamefont{and} \bibinfo{author}{\bibfnamefont{K.~W.} \bibnamefont{West}},
  \bibinfo{journal}{Phys. Rev. B} \textbf{\bibinfo{volume}{79}},
  \bibinfo{pages}{161308(R)} (\bibinfo{year}{2009}{\natexlab{b}}).

\bibitem[{\citenamefont{Hatke et~al.}(2011{\natexlab{e}})\citenamefont{Hatke,
  Zudov, Pfeiffer, and West}}]{hatke:2011a}
\bibinfo{author}{\bibfnamefont{A.~T.} \bibnamefont{Hatke}},
  \bibinfo{author}{\bibfnamefont{M.~A.} \bibnamefont{Zudov}},
  \bibinfo{author}{\bibfnamefont{L.~N.} \bibnamefont{Pfeiffer}},
  \bibnamefont{and} \bibinfo{author}{\bibfnamefont{K.~W.} \bibnamefont{West}},
  \bibinfo{journal}{Phys. Rev. B} \textbf{\bibinfo{volume}{83}},
  \bibinfo{pages}{081301(R)} (\bibinfo{year}{2011}{\natexlab{e}}).

\bibitem[{not({\natexlab{b}})}]{note:2}
\bibinfo{note}{If $\Delta \propto B_\parallel$, then the argument of the Dingle
  factor would acquire a correction $-\pi\Delta/\oc \propto
  -B_\parallel/B_\perp = -\tan \theta$, which is independent of the oscillation
  order.}

\bibitem[{\citenamefont{Romero et~al.}(2008)\citenamefont{Romero, McHugh,
  Sarachik, Vitkalov, and Bykov}}]{romero:2008}
\bibinfo{author}{\bibfnamefont{N.}~\bibnamefont{Romero}},
  \bibinfo{author}{\bibfnamefont{S.}~\bibnamefont{McHugh}},
  \bibinfo{author}{\bibfnamefont{M.~P.} \bibnamefont{Sarachik}},
  \bibinfo{author}{\bibfnamefont{S.~A.} \bibnamefont{Vitkalov}},
  \bibnamefont{and} \bibinfo{author}{\bibfnamefont{A.~A.} \bibnamefont{Bykov}},
  \bibinfo{journal}{Phys. Rev. B} \textbf{\bibinfo{volume}{78}},
  \bibinfo{pages}{153311} (\bibinfo{year}{2008}).

\end{thebibliography}

\end{document}